\documentclass[twoside,a4paper,12pt]{article}
\usepackage[utf8]{inputenc}
\usepackage[T1]{fontenc}
\usepackage{textcomp}
\usepackage{lmodern}
\usepackage[english]{babel}
\usepackage{amsmath}
\usepackage{amsfonts}
\usepackage{amsthm}
\usepackage{amssymb}
\usepackage{layout}
\usepackage[top=2.5cm, bottom=2.5cm, left=2.5cm, right=2.5cm]{geometry}
\usepackage{graphicx}
\usepackage{color}
\usepackage{xcolor}
\usepackage{xfrac}
\usepackage{bm}
\usepackage[extra]{tipa}
\usepackage{ulem}
\usepackage{fancyhdr}
\usepackage{wrapfig}
\usepackage{epstopdf}
\usepackage{float}
\usepackage{authblk}
\usepackage{hyperref}
\usepackage{authblk}
\usepackage[toc,page]{appendix}
\usepackage{chemist} 
\usepackage[version=3]{mhchem} 
\usepackage{subfigure}
\usepackage{caption}

\title{Stochastic approach to Fisher and Kolmogorov, Petrovskii, and Piskunov wave fronts for species with different diffusivities in dilute and concentrated solutions}

\author[1,2]{Gabriel Morgado}
\author[1]{Bogdan Nowakowski}
\author[2]{Annie Lemarchand*}
\affil[1]{Institute of Physical Chemistry, Polish Academy of Sciences, Kasprzaka 44/52, 01-224 Warsaw, Poland}
\affil[2]{Laboratoire de Physique Th\'eorique de la Mati\`ere Condens\'ee, Sorbonne Universit\'e, CNRS UMR 7600, 4 place Jussieu, case courrier 121, 75252 Paris CEDEX 05, France}

\begin{document}
\maketitle
* Corresponding author: Annie Lemarchand, E-mail: annie.lemarchand@sorbonne-universite.fr \\


\begin{abstract}
A wave front of Fisher and Kolmogorov, Petrovskii, and Piskunov type involving
two species A and B with different diffusion coefficients $D_A$ and $D_B$ is studied using a master equation approach
in dilute and concentrated solutions. 
Species A and B are supposed to be engaged in the autocatalytic reaction A+B $\rightarrow$ 2A.
Contrary to the results of a deterministic description, the front speed deduced from the master equation in the dilute case
sensitively depends on the diffusion coefficient of species B. A linear analysis of the deterministic equations
with a cutoff in the reactive term cannot explain the decrease of the front speed observed for $D_B>D_A$. 
In the case of a concentrated solution, the transition rates associated with cross-diffusion are derived from the 
corresponding diffusion fluxes.
The properties of the wave front obtained in the dilute case 
remain valid but are mitigated by cross-diffusion which reduces the impact of different diffusion coefficients. 
\end{abstract}

\newpage

\baselineskip=24pt
\section{Introduction}
Wave fronts propagating into an unstable state according to the model of
Fisher and Kolmogorov, Petrovskii, and Piskunov (FKPP)~\cite{fisher,kpp}
are encountered in many fields~\cite{vansaarloos}, in particular biology~\cite{murray} 
and ecology~\cite{mendez14}. Phenotype selection through the propagation of the fittest trait \cite{bouin} and
cultural transmission in neolithic transitions \cite{fort2016} are a few examples of applications of FKPP fronts.
The model introduces a partial differential equation with a logistic growth term and a diffusion term.
\\

The effect of non standard diffusion on the speed of FKPP front is currently investigated~\cite{mancinelli,froemberg,cabre,adnani} 
and we recently considered the propagation of a wave front in a concentrated solution in which cross-diffusion cannot be neglected~\cite{pre19}.
Experimental evidence of cross-diffusion has been given in systems involving ions, micelles, surface, 
or polymer reactions and its implication in hydrodynamic instabilities has been demonstrated~\cite{vanag1,leaist,vanag2,rossi,budroni1,budroni2}.
In parallel, cross-diffusion is becoming an active field of research in applied mathematics~\cite{desvillettes1,desvillettes2,desvillettes3,juengel,daus,moussa}.

The sensitivity of FKPP fronts to fluctuations has been first numerically observed~\cite{breuer94,lemarchand95}.
An interpretation has been then proposed in the framework of a deterministic approach introducing a cutoff in the logistic term~\cite{brunet97}.
In mesoscopic or microscopic descriptions of the invasion front of A particles
engaged in the reaction $\rm{A}+\rm{B} \rightarrow 2\rm{A}$,
the discontinuity induced by the rightmost particle in the leading edge of species A profile amounts to a cutoff in the reactive term.
The inverse of the number of particles in the reactive interface gives an estimate of the cutoff~\cite{hansen}.
The study of the effect of fluctuations on FKPP fronts remains topical~\cite{panja2,doering}. 
In this paper we perform a stochastic analysis of a reaction-diffusion front of FKPP type in the case of two species A and B with 
different diffusion coefficients~\cite{mai}, giving rise to cross-diffusion phenomena in concentrated solutions.

The paper is organized as follows. Section 2 is devoted to a dilute system without cross-diffusion. The effects of the discrete number of particles on the front speed, 
the shift between the profiles of the two species and the width of species A profile are deduced from a master equation approach.
In section 3, we derive the expression of the master equation associated with a concentrated system inducing cross-diffusion
and compare the properties of the FKPP wave front in the dilute and the concentrated cases.
Conclusions are given in section 4.

\section{Dilute system}
We consider two chemical species A and B engaged in the reaction
\begin{equation}
    \ce{A + B ->[k] 2 A},
    \label{reac}
\end{equation}
where $k$ is the rate constant. The diffusion coefficient, $D_A$, of species A may differ from the diffusion coefficient, $D_B$, of species B.  

In a deterministic approach, the reaction-diffusion equations are
\begin{eqnarray}
    \partial_t A &=& D_A\partial_x^2A + kAB \label{RDA}\\
     \partial_t B &=& D_B\partial_x^2B - kAB \label{RDB}
\end{eqnarray}
where the concentrations of species A and B are denoted by $A$ and $B$. 
The system admits wave front solutions propagating without deformation at constant speed. 
For sufficiently steep initial conditions and in particular step functions $(A(x,t=0)=C_0H(-x)$ and $B(x,t=0)=C_0H(x))$, where
$C_0$ is constant and $H(x)$ is the Heaviside function, the minimum velocity 
\begin{eqnarray}
v^*=2\sqrt{kC_0D_A}
\label{vdet}
\end{eqnarray}
is selected~\cite{vansaarloos,murray,brunet97}.
The parameter $C_0=A(x,0)+B(x,0)$ is the sum of the initial concentrations of species A and B. 
Discrete variables of space, $i=x/\Delta x$, and time, $s=t/\Delta t$, where $\Delta x$ is the cell length and $\Delta t$ is the time step, are introduced in order to numerically solve Eqs. (\ref{RDA}) and (\ref{RDB}) in a wide range of diffusion coefficients $D_B$. We consider a system of $\ell=2000$ spatial cells.
The initial condition is a step function located in the cell $i_0=\ell/2$
\begin{eqnarray}
A(i,0)&=&C_0H(i_0-i),\\
B(i,0)&=&C_0H(i-i_0),
\end{eqnarray}
where $H(i)$ is the Heaviside function.
In order to simulate a moving frame and to counterbalance the autocatalytic production of species A in a finite system, 
the following procedure is applied.
At the time steps $s$ such that $\sum_{i=1}^{\ell} A(i,s) > \sum_{i=1}^{\ell} A(i,0)$,
the first cell is suppressed and a last cell with $A(\ell,s)=0$ and $B(\ell,s)=C_0$ is created.
Hence, the inflection point of the front profile remains close to the initial step of the Heaviside function.

In small systems with typically hundreds of particles per spatial cell, the deterministic description may fail and a stochastic approach is required.
We consider the chemical master equation associated with Eq.~(\ref{reac})~\cite{nicolis,gardiner}. The master equation is divided into two parts
\begin{eqnarray}
\label{med}
\dfrac{\partial P(\phi)}{\partial t}=\left.\dfrac{\partial P(\phi)}{\partial t}\right|_{\rm reaction}+\left.\dfrac{\partial P(\phi)}{\partial t}\right|_{\rm diffusion}
\end{eqnarray}
where the first part corresponds to the reactive terms
\begin{eqnarray}
\label{medr}
\left.\dfrac{\partial P(\phi)}{\partial t}\right|_{\rm reac}=&\sum_i\dfrac{k}{\Omega N_0}\bigg[(N_A(i)-1)(N_B(i)+1)P(\{N_A(i)-1,N_B(i)+1\}) \nonumber \\
 & -N_A(i)N_B(i)P(\phi)\bigg]
\end{eqnarray}
and the second part corresponds to the diffusion terms
\begin{eqnarray}
\label{medd}
\left.\dfrac{\partial P(\phi)}{\partial t}\right|_{\rm diff}=&\sum_i\bigg[\dfrac{D_A}{\Delta x^2}(N_A(i)+1)\big[P(\{N_A(i-1)-1,N_A(i)+1\}) \nonumber \\
&+P(\{N_A(i)+1,N_A(i+1)-1\})\big] \nonumber \\
&+\dfrac{D_B}{\Delta x^2}(N_B(i)+1)\big[P(\{N_B(i-1)-1,N_B(i)+1\})\nonumber \\
&+P(\{N_B(i)+1,N_B(i+1)-1\})\big] \nonumber\\
&-\dfrac{2}{\Delta x^2}\big(D_AN_A(i)+D_BN_B(i)\big)P(\phi)\bigg]
\end{eqnarray}
where $\phi=\{N_A(i),N_B(i)\}$ denotes the default state, $\Omega$, the typical size of the system, $N_0=\Omega C_0$, the initial total number of particles in a cell, and
$N_A(i)=\Omega A(i)$ and $N_B(i)=\Omega B(i)$ are the numbers of particles A and B in cell $i$.
We consider parameter values leading to the macroscopic values used in the deterministic approach. 
The initial condition is given by $(N_A(i)=N_0,N_B(i)=0)$ for $1 \leq i < \ell/2$
and $(N_A(i)=0,N_B(i)=N_0)$ for $\ell/2 \leq i \leq \ell$ with $N_0=100$, $\Omega=10$ $(C_0=10)$.

The kinetic Monte Carlo algorithm developed by Gillespie is used to directly simulate the reaction and diffusion processes and numerically solve the master equation~\cite{gillespie}. 
The procedure used in the deterministic approach to evaluate the front speed is straightforwardly extended to the fluctuating system.

\begin{figure}
\centering
\includegraphics[height=6cm]{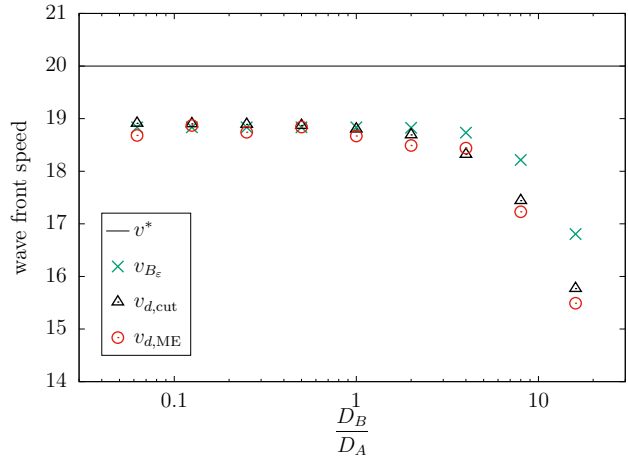}
\caption{Dilute system. Wave front speeds $v_{d,{\rm ME}}$, $v_{d,{\rm cut}}$, $v_{B_\varepsilon}$, and $v_d=v^*$ versus ratio of diffusion coefficients $D_B/D_A$ in the dilute case. 
The values of $v_{d,{\rm ME}}$ (red circles) 
are deduced from the direct simulation of the master equation (Eqs. (\ref{med}-\ref{medd})) for $k=10$, $\Omega=10$, $N_0=100$, $D_A=1$, $\ell=2000$, and $\Delta x=0.008$. 
The values of $v_{d,{\rm cut}}$ (black open triangles) are deduced from the numerical integration of the deterministic equations (Eqs.~(\ref{RDAc}) and (\ref{RDBc})) in the presence of a cutoff $\varepsilon=10^{-4}$ for $k=10$, $C_0=10$, $D_A=1$, $\ell=2000$, $\Delta x=0.008$, and $\Delta t= 6.4\times 10^{-6}$.
The values of $v_{B_\varepsilon}$ (green crosses) are deduced from Eq.~(\ref{vbeps}) 
in which the value $B_\varepsilon$ has been deduced from the numerical integration of Eqs.~(\ref{RDAc}) and (\ref{RDBc}).
The horizontal line gives the minimum velocity $v_d=v^*$ (Eq. (\ref{vdet})) of an FKPP front in the absence of a cutoff.
}
\label{figv}
\end{figure}

\subsection{Front speed}
For sufficiently small spatial lengths $\Delta x$ and time steps $\Delta t$,
the numerical solution of the deterministic equations given in Eqs. (\ref{RDA}) and (\ref{RDB}) leads to the same propagation speed $v_d$, 
where the index $d$ stands for dilute, in the entire range of $D_B/D_A$ values~\cite{pre19}. 
The number of cells created during $10^7$ time steps once a stationary propagation is reached is used to evaluate the front speed.
For the chosen parameter values, we find a propagation speed obeying $v_d=v^*=20$ 
with an accuracy of $0.4\%$: No appreciable deviation from the unperturbed deterministic prediction given in Eq. (\ref{vdet}) is observed.
In particular, the front speed $v_d$ does not depend on the diffusion coefficient $D_B$.
The front speed deduced from the direct simulation of Eqs. (\ref{med}-\ref{medd}) is denoted $v_{d,{\rm ME}}$ where the index $d$ stands for dilute and the index ${\rm ME}$ for master equation.
As shown in Fig.~\ref{figv}, the velocity $v_{d,{\rm ME}}$ is smaller than the deterministic prediction $v^*$ 
given in Eq.~(\ref{vdet}).

As long as $D_B$ remains smaller than or equal to $D_A$, the velocity $v_{d,{\rm ME}}$ is constant. 
The main result of the master equation approach is that the front speed drops 
as $D_B$ increases above $D_A$. Typically, for $D_B/D_A=16$, the velocity $v_{d,{\rm ME}}$
is reduced by $22\%$ with respect to $v_d=v^*$. Due to computational costs, larger $D_B/D_A$ values were not investigated.

In the case of identical diffusion coefficients for the two species, the decrease of the front speed observed in a stochastic description is interpreted in the framework of the cutoff approach introduced by Brunet and Derrida~\cite{brunet97}. For $D_A=D_B$, the dynamics of the system is described by a single equation. When a cutoff $\varepsilon$ is introduced in the reactive term according to
\begin{equation}
\partial_tA=\partial_x^2A + kA(C_0-A)H(A-\varepsilon),
\end{equation} 
the velocity is given by 
\begin{eqnarray}
v_\varepsilon=v^*\left(1-\dfrac{\pi^2}{2(\ln \varepsilon)^2}\right)
\label{veps}
\end{eqnarray}
In a particle description, the cutoff is interpreted as the inverse of the total number of particles in the reactive interface~\cite{hansen}: 
\begin{equation}
\label{cutoff}
\varepsilon=\dfrac{\Delta x}{N_0W^*}
\end{equation}
where the width of the interface is roughly evaluated at~\cite{murray,pre19}
\begin{equation}
W^*=8\sqrt{\dfrac{D_A}{kC_0}}
\end{equation}
For the chosen parameter values, the cutoff equals $\varepsilon=10^{-4}$ leading to the corrected speed $v_\varepsilon=18.84$. 
According to Fig.~\ref{figv}, the velocity $v_{d,{\rm ME}}$ deduced from the master equation for $D_A=D_B$ agree with the velocity
$v_\varepsilon$ deduced from the cutoff approach. The results are unchanged for $D_B<D_A$ and Eq.~(\ref{veps}) 
correctly predicts the velocity in a fluctuating system. For $D_B>D_A$, Eq.~(\ref{veps}) is not valid. 
Nevertheless, the relevance of the cutoff approach can be checked by numerically integrating the two following equations 
\begin{eqnarray}
     \partial_t A &= D_A\partial_x^2A + kABH(A-\varepsilon)\label{RDAc}\\
     \partial_t B &= D_B\partial_x^2B - kABH(A-\varepsilon)\label{RDBc}
\end{eqnarray}
The values of the front speed $v_{d,{\rm cut}}$ deduced from the numerical integration
of Eqs. (\ref{RDAc}) and (\ref{RDBc}) are given in Fig.~\ref{figv} and satisfactorily agree with the results $v_{d,{\rm ME}}$ of the master equation, including for large $D_B/D_A$ values. 

\begin{figure}
\centering
\subfigure{\includegraphics[height=6cm]{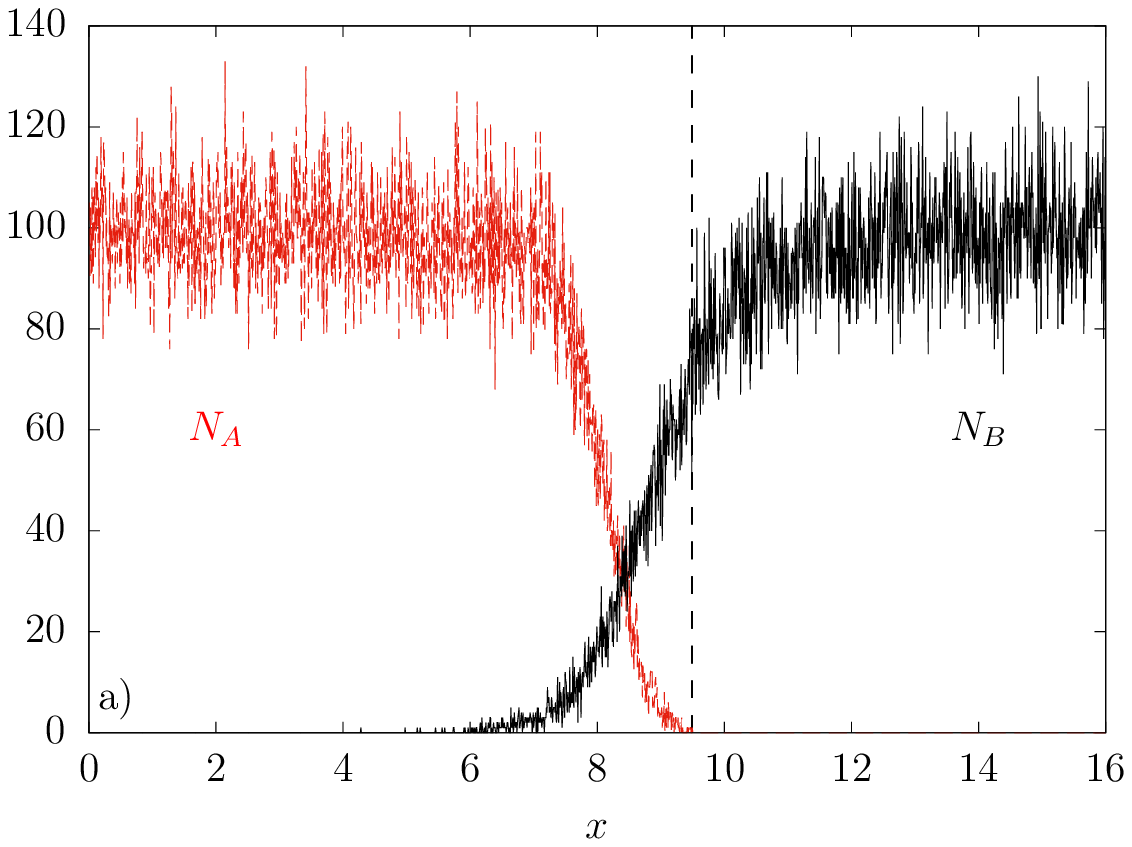}}
\subfigure{\includegraphics[height=6cm]{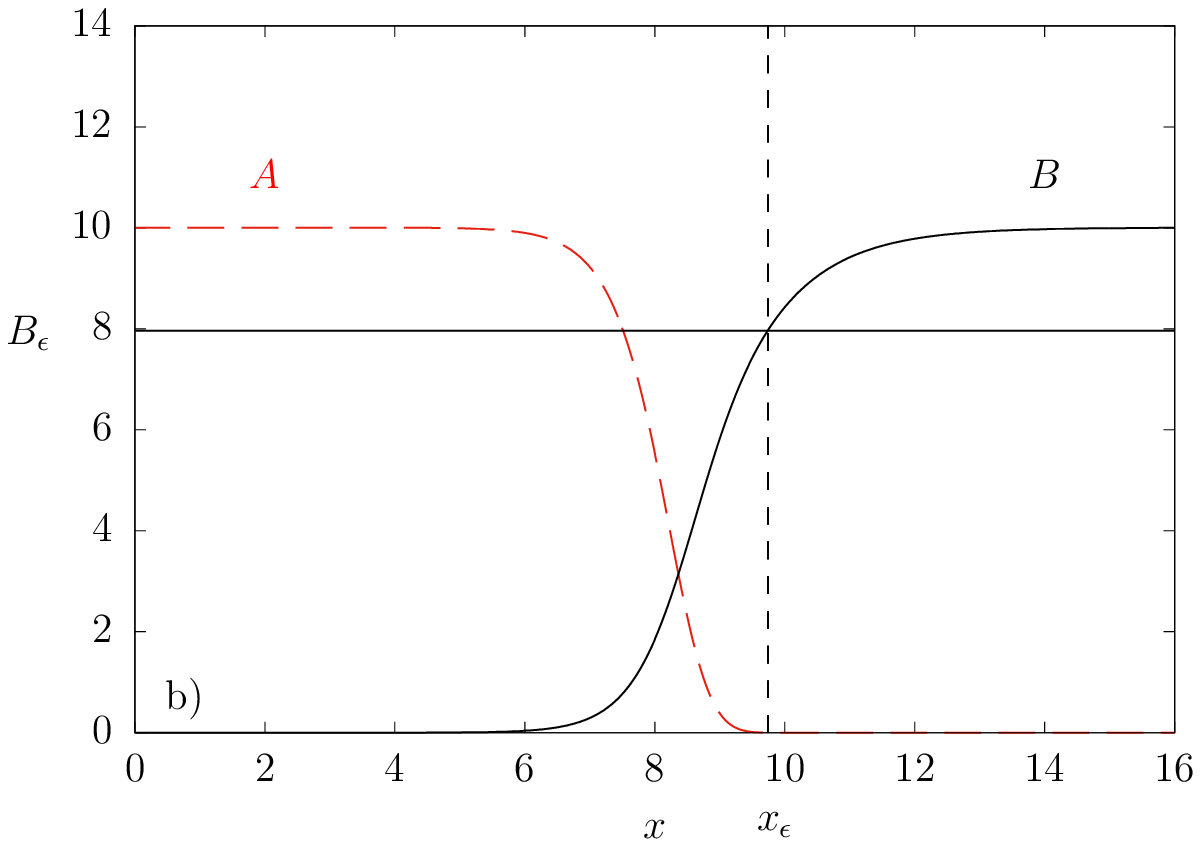}}
\caption{Dilute system. (a) Numbers $N_A$ of particles A (red dashed line) and $N_B$ of particles B (black solid line) 
versus spatial coordinate $x$ deduced from direct simulation of
the master equation (Eqs. (\ref{med}-\ref{medd})) using Gillespie method. The snapshot is given at time $t=9$ for $k=10$, $\Omega=10$, $N_0=100$, $D_A=1$, $D_B=16$, $\ell=2000$, and $\Delta x=0.008$.
The vertical dashed line indicates the rightmost cell occupied by A particles. (b) Concentrations $A$ of species A (red dashed line) and $B$ of 
species B (black solid line) versus spatial 
coordinate $x$ deduced from numerical integration of the deterministic equations (Eqs.~(\ref{RDAc}) and (\ref{RDBc})) in the presence of a cutoff $\varepsilon=10^{-4}$ . The snapshot is given at time $t=640$ for the same other parameters as in the master 
equation approach. The vertical dashed line indicates the abscissa $x_\varepsilon$ for which the scaled A concentration $A(x_\varepsilon)/C_0$ reaches the cutoff value.
 The horizontal line indicates the value $B_\varepsilon$ of B concentration at the abscissa $x_\varepsilon$.}
\label{prof}
\end{figure}

\begin{figure}
\centering
\includegraphics[height=6cm]{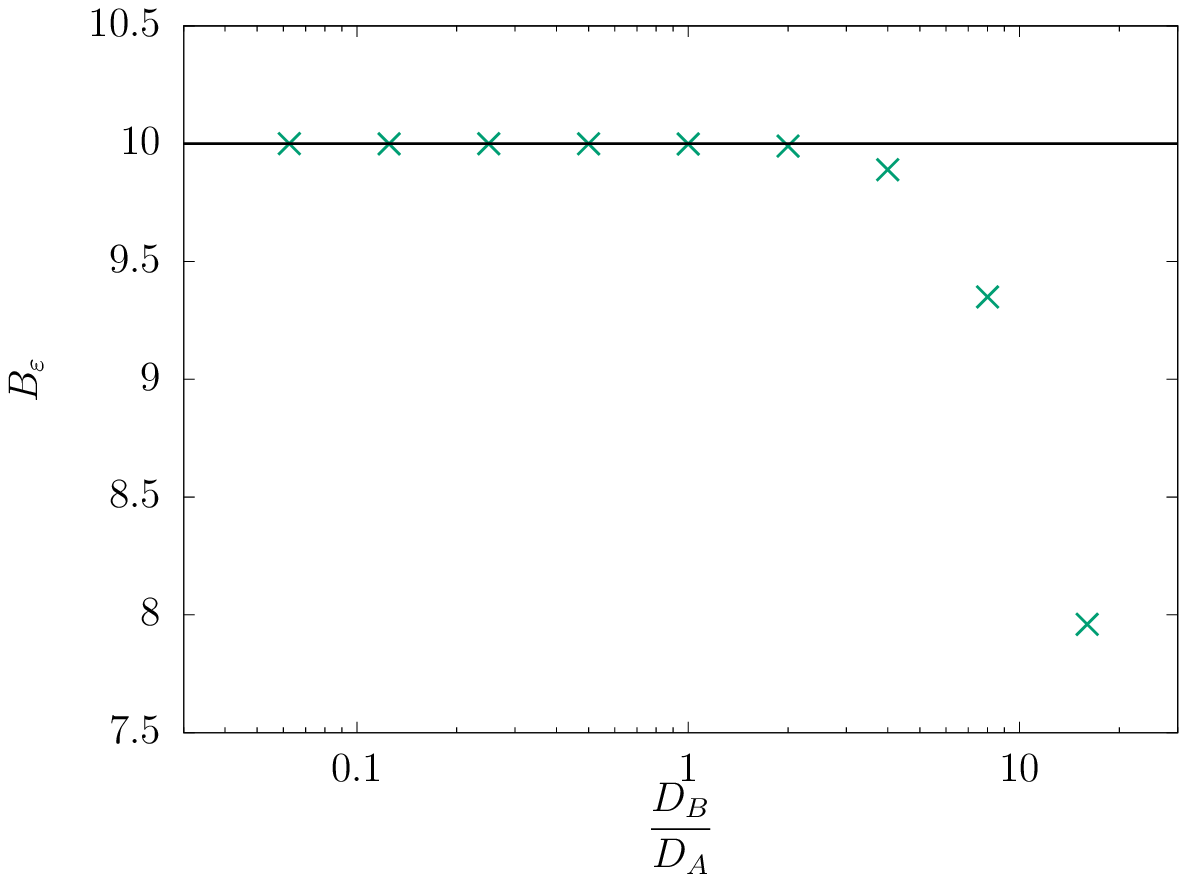}
\caption{Dilute system. The green crosses give the value $B_\varepsilon$ deduced from the numerical integration of the deterministic equations (Eqs.~(\ref{RDAc}) and (\ref{RDBc})) with a cutoff $\varepsilon=10^{-4}$
  versus the ratio of the diffusion coefficients $D_B/D_A$. The horizontal line indicates the concentration $C_0$.
The parameters are given in the caption of Fig.~\ref{figv}.}
\label{beps}
\end{figure}

According to Fig.~\ref{prof}a, the A profile is steeper than the B profile for $D_B>D_A$. The mean number of B particles in the leading edge smoothly converges to $N_0$. In average, the rightmost A particle sees a number of B particles smaller than $N_0$.
The significant decrease of the front velocity $v_{d,{\rm cut}}$ for $D_B>D_A$ is qualitatively interpreted by the apparent number $N_\varepsilon$ of B particles seen by the rightmost A particle in the leading edge.
The linear analysis of Eqs.~(\ref{RDAc}) and (\ref{RDBc}) according to the cutoff approach~\cite{brunet97} leads to Eq.~(\ref{veps}) 
which does not account for the behavior at large $D_B$. A nonlinear analysis would be necessary. Using the perturbative approach 
that we developed in the case of the deterministic description~\cite{murray,pre19}, applying the Hamilton-Jacobi 
technique~\cite{fedotov1999,mirrahimi}, or deducing the variance $\langle AB\rangle$ from a Langevin approach~\cite{carlo}, 
we unsuccessfully tried to find an analytical estimation of the front speed. 
Instead, we suggest the following empirical expression of the velocity of an FKPP front for two species with different diffusion coefficients
\begin{equation}
v_{B_\varepsilon}=2\sqrt{kB_\varepsilon D_A}\left(1-\dfrac{\pi^2}{2(\ln\varepsilon)^2}\right)
\label{vbeps}
\end{equation}
where $B_\varepsilon$ denotes the concentration of B species at the abscissa $x_\varepsilon$ at which the scaled 
concentration $A(x_\varepsilon)/C_0$ is equal to the cutoff $\varepsilon$ (see Fig.~\ref{prof}b).
The variation of $B_\varepsilon$ versus $D_B/D_A$ is numerically evaluated using Eqs.~(\ref{RDAc}) and (\ref{RDBc}). 
The result is given in Fig.~\ref{beps}. 

As shown in Fig.~\ref{figv}, the variation of the front speed $v_{B_\varepsilon}$ with $D_B/D_A$ deduced from Eq. (\ref{vbeps}) slightly underestimates the results 
$v_{d,{\rm cut}}$ deduced from the numerical integration of the deterministic equations (Eqs.~(\ref{RDAc}) and (\ref{RDBc})) with a cutoff.

\subsection{Profile properties}
We focus on two steady properties of the wave front, the shift between the profiles of species A and B 
and the width of species A profile~\cite{pre19}.

\begin{figure}
\centering
\includegraphics[height=6cm]{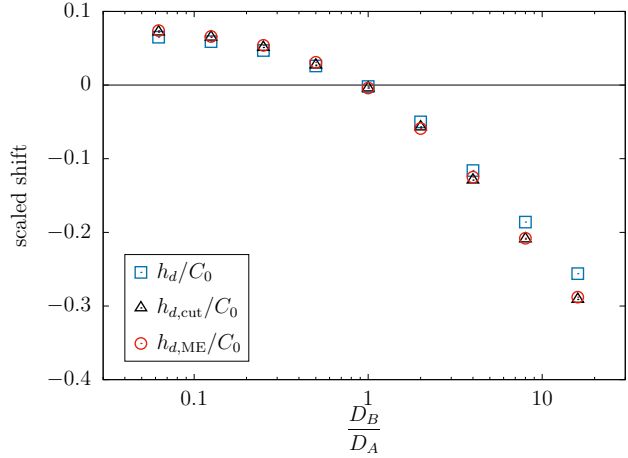}
\caption{Dilute system. Scaled shifts $h_{d,{\rm ME}}/C_0$, $h_{d,{\rm cut}}/C_0$, and $h_d/C_0$ between the profiles of species A and B versus ratio of diffusion coefficients $D_B/D_A$.
The values of $h_{d,{\rm ME}}/C_0$ (red circles) are deduced from the master equation (Eqs. (\ref{med}-\ref{medd})).
The values of $h_{d,{\rm cut}}/C_0$ (black open triangles) are deduced from the deterministic equations (Eqs. (\ref{RDAc} and \ref{RDBc})) 
with a cutoff $\varepsilon=10^{-4}$.
The values of $h_d/C_0$ (blue open squares) are deduced from the deterministic equations (Eqs. (\ref{RDA}) and (\ref{RDB})) without cutoff.
The line gives the results for $D_A=D_B$.
The parameters are given in the caption of Fig. 1.}
\label{hd}
\end{figure}

For a wave front propagating at speed $v$ and using the coordinate $z=x-vt$ in the moving frame, the shift between the profiles of the two species 
is defined as the difference $A(z=0)-B(z=0)$ of concentrations between species A and B at the origin
$z=0$ chosen such that $A(z=0)=C_0/2$. 
The shift is denoted by $h_d$, where the index $d$ stands for dilute, when the concentrations are solutions of the deterministic equations
without cutoff given in Eqs. (\ref{RDA}) and (\ref{RDB}).
As shown in Fig.~\ref{hd}, the shift $h_d$ significantly varies with the ratio $D_B/D_A$, in particular when $D_B$ is larger than
$D_A$ \cite{pre19}. The shift vanishes for $D_A=D_B$, is positive
for $D_B<D_A$ and negative for $D_B>D_A$.

The direct simulation of the master equation leads to highly fluctuating profiles. We use the following strategy to compute the shift $h_{d,{\rm ME}}$.
First, starting from the leftmost cell, we scan to the right to determine the label $i_l$ of the first cell in which the number of A particles drops under $N_0/2$ and store
$N_B(i_l,s)$ for a large discrete time $s$ at which the profile has reached a steady shape.
Then, starting from the rightmost cell labeled $\ell$, we follow a similar procedure and determine the label $i_r$ of the first cell in which the number
of A particles overcomes $N_0/2$ and store $N_B(i_r,s)$ for the same discrete time $s$. The instantaneous value of the shift deduced from the master equation at discrete time $s$ is then given by
$(N_0-N_B(i_l,s)-N_B(i_r,s))/2\Omega$. The values of the shift $h_{d,{\rm ME}}$ used to draw Fig.~\ref{hd} are obtained after a time average between the times $t=1$ and $t=10$ in arbitrary units, i.e.
between $s=1.5\times 10^5$ and $s=1.5\times 10^6$ in number of time steps. 

The shift $h_{d,{\rm ME}}$ between the profiles of A and B is sensitive to the fluctuations
of the number of particles described by the master equation.  
Introducing an appropriate cutoff satisfying Eq. (\ref{cutoff}) in the reactive term of
the deterministic equations given in Eqs. (\ref{RDAc}) and (\ref{RDBc}) leads to values of the shift $h_{d,{\rm cut}}$ in very good agreement with the results 
$h_{d,{\rm ME}}$ of the master equation.

\begin{figure}
\centering
\includegraphics[height=6cm]{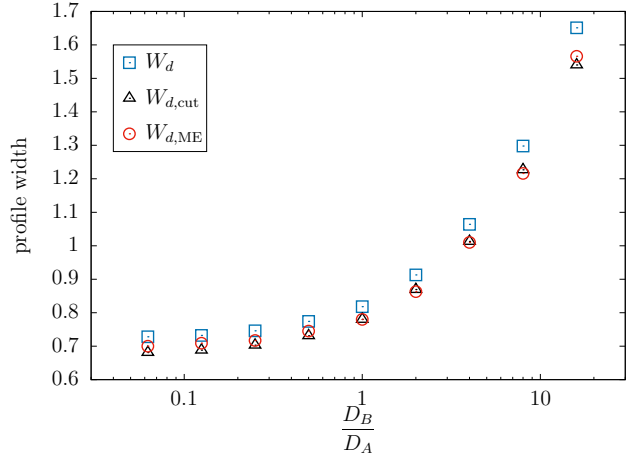}
\caption{Dilute system. Profile widths deduced from different approaches versus ratio of diffusion coefficients $D_B/D_A$.
The values of $W_{d,{\rm ME}}$ (red circles) are deduced from the master equation (Eqs. (\ref{med}-\ref{medd})).
The values of $W_{d,{\rm cut}}$ (black open triangles) are deduced from the numerical integration of the deterministic equations (Eqs. (\ref{RDAc}) and (\ref{RDBc})) with a cutoff $\varepsilon=10^{-4}$.
The values of $W_d$ (blue open squares) are deduced from the numerical integration of the deterministic equations (Eqs. (\ref{RDA}) and (\ref{RDB}))
without cutoff.
The parameters are given in the caption of Fig. 1.}
\label{Wd}
\end{figure}

Considering the deterministic equations, we deduce the width of A profile from the steepness $A'(0)$ in the moving frame at the origin $z=0$ and find 
\begin{equation}
W_d=C_0/|A'(0)|
\label{width}
\end{equation}
where $A$ is solution of Eqs. (\ref{RDA}) and (\ref{RDB}) without cutoff.
The same definition is applied to Eqs. (\ref{RDAc}) and (\ref{RDBc}) to obtain the width $W_{d,{\rm cut}}$ in the presence of a cutoff.
The definition has to be adapted to take into account the fluctuations of the profile deduced from the master equation.
Using the cell labels $i_l$ and $i_r$ determined for the shift between the fluctuating A and B profiles solutions of Eqs. (\ref{med}-\ref{medd}), 
we define the mean cell label $i_m$ as the nearest integer to the average $(i_l+i_r)/2$.
We use Eq. (\ref{width}) with $|A'(0)| \simeq (N_A(i_m-40)-N_A(i_m+40))/(81\Delta x\Omega)$
to compute the instantaneous width. As in the case of the shift $h_{d,{\rm ME}}$ between the fluctuating profiles of A and B, the values $W_{d,{\rm ME}}$ of the width used to draw Fig.~\ref{Wd}
are obtained after a time average between the times $t=1$ and $t=10$.

As shown in Fig.~\ref{Wd}, the width $W_d$ deduced from the deterministic equations without cutoff is smaller (resp. larger) for $D_B<D_A$ (resp. $D_B>D_A$)
than the width evaluated at $W^*$ in the case of identical diffusion coefficients $D_B=D_A$ \cite{pre19}.  
The width $W_{d,{\rm ME}}$ deduced from the master equation (Eqs. (\ref{med}-\ref{medd})) and 
the width $W_{d,{\rm cut}}$ deduced from the deterministic equations (Eqs. (\ref{RDAc}) and (\ref{RDBc})) with a cutoff obeying Eq. (\ref{cutoff}) agree and are both smaller than the width $W_d$
of the wave front, solution of the deterministic equations without cutoff.

According to the good agreement between the results of the master equation and the deterministic equations with a cutoff,
it is more relevant to describe the effect of the fluctuations on the wave front as the effect of the discretization of the variables than a pure noise effect.

\begin{figure}
\centering
\includegraphics[height=6cm]{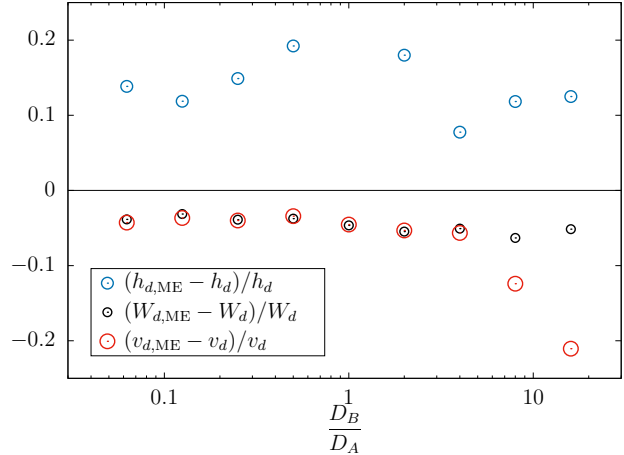}
\caption{Dilute system. Relative differences between 
the front properties deduced from the master equation (Eqs. (\ref{med}-\ref{medd})) and
the analogous properties deduced from the deterministic equations without cutoff (Eqs. (\ref{RDA} and (\ref{RDB}))
versus $D_B/D_A$. The large red circles give the relative difference $(v_{d,{\rm ME}}-v_d)/v_d$ for the front speed,
the blue circles of intermediate size give the relative difference $(h_{d,{\rm ME}}-h_d)/h_d$ for the shift between A and B profiles, and
the small black circles give the relative difference $(W_{d,{\rm ME}}-W_d)/W_d$ for the width of A profile.
The parameters are given in the caption of Fig. 1.}
\label{figXX}
\end{figure}

Figure \ref{figXX} summarizes the effect of the fluctuations on the three quantities $q$ for $q=v,h,W$
in the whole range of considered values of the ratio $D_B/D_A$ for the dilute system.
The relative differences $(q_{d,{\rm ME}}-q_d)/q_d$ between the results deduced from the master equation and the deterministic equations without cutoff
are given in Fig. \ref{figXX} for the velocity, the shift, and the width.
In the whole range of $D_B/D_A$, the discrete nature of the number of particles in the master equation 
induces a small decrease of $5\%$ of the profile width with respect to the deterministic description without cutoff.
A significant increase of $14\%$ of the shift between the A and B profiles is observed in the presence of fluctuations in the entire interval
of ratios of diffusion coefficients. 
As for the width, the relative difference of velocity $(v_{d,{\rm ME}}-v_d)/v_d$, with $v_d=v^*$, is negative and takes the same value
of $-5\%$ for $D_B/D_A \leq 1$. However, the relative difference of velocity is not constant for $D_B/D_A > 1$ and reaches $-22\%$ for
$D_B/D_A=16$.
Hence, a significant speed decrease is observed whereas the shift and the width, 
far behind the leading edge of the front, are not affected by large diffusion coefficients of species B with respect to the
diffusion coefficient of species A. 

\section{Concentrated system}
In a dilute system, the solvent S is in great excess with respect to the reactive species A and B.
The concentration of the solvent is then supposed to remain homogeneous regardless of the 
variation of concentrations $A$ and $B$.
In a concentrated solution, the variation of the concentration of the solvent cannot be ignored.
In the linear domain of irreversible thermodynamics, the diffusion fluxes are linear combinations of the concentration gradients of the different species. The flux $j_X$ of species X=A, B, S depends on the concentration gradients and the diffusion coefficients of all species A, B, and S~\cite{degroot,signon16}. Using the conservation relations $C_{tot}=A+B+S$, where $C_{tot}$ is a constant, 
we eliminate the explicit dependence of the fluxes on the concentration $S$ of the solvent and find
\begin{eqnarray}
j_A&=&-\left(1-\dfrac{A}{C_{tot}}\right)D_A\partial_xA+\dfrac{A}{C_{tot}}D_B\partial_xB \label{ja}\\
j_B&=&\dfrac{B}{C_{tot}}D_A\partial_xA-\left(1-\dfrac{B}{C_{tot}}\right)D_B\partial_xB \label{jb}
\end{eqnarray}
According to the expression of the diffusion fluxes in a concentrated system, the reaction-diffusion equations associated with the chemical mechanism given in Eq.~(\ref{reac}) read~\cite{signon16}
\begin{eqnarray}
    \partial_t A &=& D_A\partial_x\left[\left(1-\dfrac{A}{C_{tot}}\right)\partial_xA\right]-D_B\partial_x\left(\dfrac{A}{C_{tot}}\partial_xB\right) + kAB \label{RDAconc}\\
     \partial_t B &=& D_B\partial_x\left[\left(1-\dfrac{B}{C_{tot}}\right)\partial_xB\right]-D_A\partial_x\left(\dfrac{B}{C_{tot}}\partial_xA\right) - kAB \label{RDBconc}
\end{eqnarray}
The discrete expression of the flux at the interface between cells $i$ and $i+1$ is related to the difference of the transition rates in the master equation
according to
\begin{eqnarray}
j_X(i+\sfrac{1}{2})&=&-\frac{1}{\Delta x}\left(T_{N_X(i+1)}^--T_{N_X(i)}^+\right)
\end{eqnarray}
where $X=A,B$, the transition rate $T_{N_X(i+1)}^-$ is associated with the jump of a particle X to the left from cell $i+1$ to cell $i$, and $T_{N_X(i)}^+$ is associated 
with the jump of a particle X to the right from cell $i$ to cell $i+1$.
Using Eqs.~(\ref{ja}) and (\ref{jb}) and replacing $\partial_xX$ by $(N_X(i+1)-N_X(i))/\Omega\Delta x$ for $X=A,B$, we assign well-chosen terms of the flux $j_X(i+\sfrac{1}{2})$ to the transition rates to the left and to the right 
\begin{eqnarray}
T_{N_A(i)}^\pm&=&\dfrac{D_A}{\Delta x^2}N_A(i)-\dfrac{N_A\left(i\pm\sfrac{1}{2}\right)}{\Omega C_{tot}\Delta x^2}\left[D_AN_A(i)-D_BN_B(i\pm 1)\right]\label{wa1}\\
T_{N_B(i)}^\pm&=&\dfrac{D_B}{\Delta x^2}N_B(i)-\dfrac{N_B\left(i\pm\sfrac{1}{2}\right)}{\Omega C_{tot}\Delta x^2}\left[D_BN_B(i)-D_AN_A(i\pm 1)\right]\label{wa2}
\end{eqnarray}
to ensure that they are positive or equal to zero for any number of particles.
A standard arithmetic mean for the number $N_X\left(i\pm\sfrac{1}{2}\right)$ of particles $X=A,B$ in the virtual cell $i\pm\sfrac{1}{2}$ cannot be used since it may lead to a non-zero transition rate when the departure cell is empty. Instead, we choose the harmonic mean between the number of particles in cells $i$ and $i\pm 1$:
\begin{eqnarray}
N_X\left(i\pm\sfrac{1}{2}\right)&=&\dfrac{N_X(i)N_X(i\pm 1)}{N_X(i)+N_X(i \pm 1)}
\end{eqnarray}
which ensures that no jump of $X$ from cell $i$ to cell $i\pm 1$ occurs when the number of particles $N_X$ vanishes in cell $i$. We checked different definitions 
of the mean obeying the latter condition and found that the results are not significantly affected 
when choosing for $N_X\left(i\pm\sfrac{1}{2}\right)$ a modified arithmetic mean which vanishes if $N_X(i)=0$ and equals $(N_X(i)+N_X(i \pm 1))/2$ otherwise, 
or a geometric mean $\sqrt{N_X(i)N_X(i\pm 1)}$.\\
 
It is worth noting that, contrary to the dilute case for which the transition rate associated with the diffusion of particles X only 
depends on the number of particles X in the departure cell, the transition rate in the concentrated case also depends on the number of particles A and B in the arrival cell. 
In the case of a concentrated system, the diffusion term reads
\begin{eqnarray}
\label{mecd}
\left.\dfrac{\partial P(\phi)}{\partial t}\right|_{\rm diff}=&\sum_i \Big[T_{N_A(i)+1}^-P(\{N_A(i-1)-1,N_A(i)+1\}) \nonumber \\
&+T_{N_A(i)+1}^+P(\{N_A(i)+1,N_A(i+1)-1\}) \nonumber \\
&+T_{N_B(i)+1}^-P(\{N_B(i-1)-1,N_B(i)+1\}) \nonumber \\
&+T_{N_B(i)+1}^+P(\{N_B(i)+1,N_B(i+1)-1\}) \nonumber \\
&-\big(T_{N_A(i)}^-+T_{N_A(i)}^++T_{N_B(i)}^-+T_{N_B(i)}^+\big)P(\phi)\Big]
\end{eqnarray}
The reaction term $\left.\dfrac{\partial P(\phi)}{\partial t}\right|_{\rm reac}$ of the master equation given in Eq.~(\ref{medr}) for the dilute system is unchanged
in the case of a concentrated system.
The kinetic Monte Carlo algorithm and the initial and boundary conditions used for the dilute system are straightforwardly extended to the concentrated system. 

\begin{figure}
\centering
\includegraphics[height=6cm]{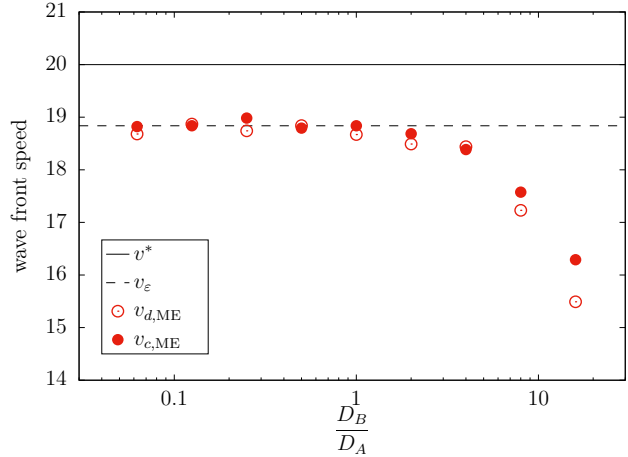}
\caption{Concentrated system. Wave front speed $v_{c,{\rm ME}}$ deduced from the master equation (Eqs. (\ref{med}), (\ref{medr}), and (\ref{mecd})) in
a concentrated system (red solid disks) for $C_{tot}=50$ and speed
$v_{d,{\rm ME}}$ deduced from the direct simulation of the master equation (Eqs. (\ref{med}-\ref{medd})) 
associated with the dilute system (red circles)
versus ratio of diffusion coefficients $D_B/D_A$. The horizontal solid line gives the minimum velocity $v^*$ (Eq. (\ref{vdet})) of an FKPP front in the absence of a cutoff.
The horizontal dashed line gives the velocity $v_\varepsilon=18.84$ given in Eq. (\ref{veps}) for 
a cutoff $\varepsilon=10^{-4}$ and $D_A=D_B$.
The parameters are given in the caption of Fig. 1.}
\label{fig4}
\end{figure}

The front speeds $v_{c,{\rm ME}}$ and $v_{d,{\rm ME}}$ deduced from the master equation in concentrated and dilute cases, respectively, are compared in Fig.~\ref{fig4}. The correction to the wave front speed induced by an increase of the ratio of diffusion coefficients $D_B/D_A$ is smaller for a concentrated system than for a   dilute system. Indeed, in the concentrated case, the diffusion of a species depends on the diffusion coefficients of both species. Hence, increasing $D_B$ at constant $D_A$ has a smaller impact on the velocity since the contribution depending on $D_B$ is partly compensated by the unchanged terms depending on $D_A$.  

\begin{figure}
\centering
\includegraphics[height=6cm]{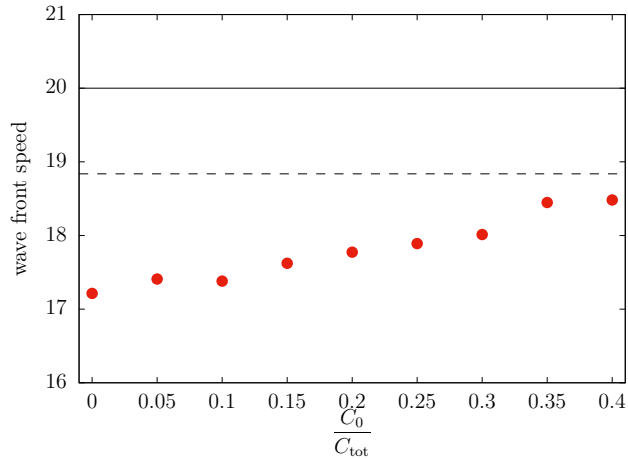}
\caption{Concentrated system. Wave front speeds versus the deviation from the dilution limit $C_0/C_{tot}$. The values of $v_{c,{\rm ME}}$ (red disks) are 
deduced from the direct simulation of the master equation (Eqs. (\ref{med}), (\ref{medr}), and (\ref{mecd})) for $k=10$, $\Omega=10$, $N_0=100$, $D_A=1$, $D_B=8$, $\ell=2000$, and $\Delta x=0.008$ ($C_0=N_0/\Omega$). 
The horizontal solid line gives the minimum velocity $v^*=20$ (Eq. (\ref{vdet})  ) 
of an FKPP front, solution of the deterministic equations (Eqs. (\ref{RDA}) and (\ref{RDB})) without cutoff.
The horizontal dashed line gives the velocity $v_\varepsilon=18.84$ given in Eq. (\ref{veps})
for a cutoff $\varepsilon=10^{-4}$ and $D_A=D_B$.}
\label{fig5}
\end{figure}

The effect of the departure from the dilution limit on the wave front speed $v_{c,{\rm ME}}$ deduced from the master equation given in 
Eqs. (\ref{med}), (\ref{medr}), and (\ref{mecd}) is shown in Fig.~\ref{fig5}. 
The dilution limit $v_{d,{\rm ME}}(D_B/D_A=8)=17.20$ 
is recovered for $C_0/C_{tot} \rightarrow 0$. 
As $C_0/C_{tot}$ increases, the solution is more concentrated and the cross-diffusion terms
become more important, so that the system is less sensitive to the difference between the diffusion coefficients $D_A$ and $D_B$: 
The wave front speed $v_{c,{\rm ME}}$ increases and tends to the value $v_\varepsilon=18.84$ predicted
by Eq. (\ref{veps}) for the cutoff $\varepsilon=10^{-4}$ and $D_A=D_B$. \\

\begin{figure}
\centering
\includegraphics[height=6cm]{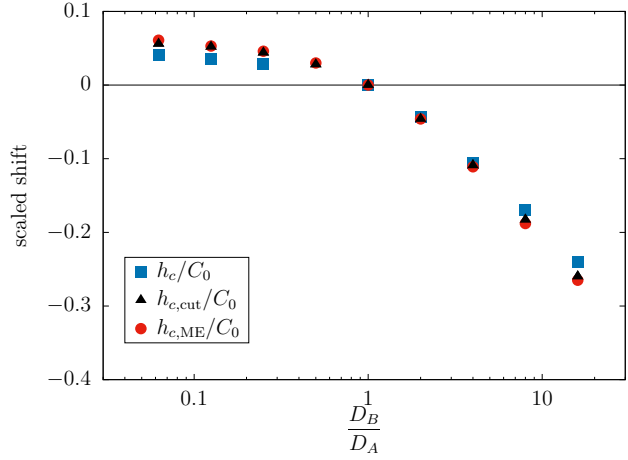}
\caption{Concentrated system. Scaled shifts $h_{c,{\rm ME}}/C_0$, $h_{c,{\rm cut}}/C_0$, and $h_c/C_0$ between the profiles of species A and B versus ratio of diffusion coefficients $D_B/D_A$.
The values of $h_{c,{\rm ME}}/C_0$ (red disks) are deduced from the master equation (Eqs. (\ref{med}), (\ref{medr}), and (\ref{mecd})).
The values of $h_{c,{\rm cut}}/C_0$ (black solid triangles) are deduced from the deterministic equations (Eqs. (\ref{RDAconc}) and (\ref{RDBconc}))
with a reactive term multiplied by the cutoff $H(A-\varepsilon)$ for $\varepsilon=10^{-4}$.
The values of $h_c/C_0$ (blue solid squares) are deduced from the deterministic equations (Eqs. (\ref{RDAconc}) and (\ref{RDBconc})) without cutoff.
The other parameters are given in the caption of Fig. \ref{fig4}.
The line gives the results for $D_A=D_B$.
}
\label{figXXX}
\end{figure}

\begin{figure}
\centering
\includegraphics[height=6cm]{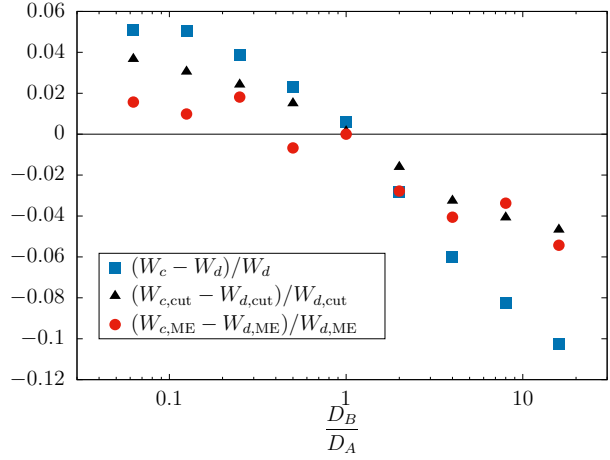}
\caption{Relative differences $(W_{c,{\rm ME}}-W_{d,{\rm ME}})/W_{d,{\rm ME}}$, $(W_{c,{\rm cut}}-W_{d,{\rm cut}})/W_{d,{\rm cut}}$, and $(W_c-W_d)/W_d$ 
between the widths in a concentrated system and a dilute system for different approaches versus $D_B/D_A$. 
The values of $W_{c,{\rm ME}}$ and $W_{d,{\rm ME}}$ (red disks) are deduced from the master equation 
(Eqs. (\ref{med}), (\ref{medr}), and (\ref{mecd}) and Eqs. (\ref{med}-\ref{medd}), respectively).
The values of $W_{c,{\rm cut}}$ and $W_{d,{\rm cut}}$ (black solid triangles) are deduced from the deterministic equations 
(Eqs. (\ref{RDAconc}) and (\ref{RDBconc}) and Eqs. (\ref{RDAc}) and (\ref{RDBc}), respectively)
with a reactive term multiplied by the cutoff $H(A-\varepsilon)$ for $\varepsilon=10^{-4}$.
The values of $W_c$ and $W_d$ (blue solid squares) are deduced from the deterministic equations 
(Eqs. (\ref{RDAconc}) and (\ref{RDBconc}) and Eqs. (\ref{RDA}) and (\ref{RDB}), respectively) without cutoff.
}
\label{figXXXX}
\end{figure}

The variation of the shifts $h_{c,{\rm ME}}$, $h_{c,{\rm cut}}$, and $h_c$ between the two profiles with respect to
the ratio of the diffusion coefficients $D_B/D_A$ 
is shown in Fig. \ref{figXXX} in a concentrated system for the three approaches, the master equation and the deterministic descriptions with and without cutoff. 
As revealed when comparing the results given in Figs.~\ref{hd} and \ref{figXXX}, the effect of the departure from the dilution limit on the shift is too small for us to evaluate
the difference $(h_{c,{\rm ME}}-h_{d,{\rm ME}})/h_{d,{\rm ME}}$ with a sufficient precision for the fluctuating results deduced from the master equations.

The effects of the departure from the dilution limit on the widths $W_{c,{\rm ME}}$, $W_{c,{\rm cut}}$, and $W_c$ of the profile are given in Fig. \ref{figXXXX} for the three approaches.
The agreement between the results $W_{c,{\rm ME}}$ and $W_{c,{\rm cut}}$ deduced from the master equation (Eqs. (\ref{med}), (\ref{medr}), and (\ref{mecd})) 
and the deterministic equations (Eqs. (\ref{RDAc} and \ref{RDBc})) with a cutoff, respectively,
is satisfying considering the high level of noise on the evaluation of the width $W_{c,{\rm ME}}$.
According to Fig.~\ref{Wd}, the width in a dilute system is smaller than the width obtained for identical diffusion coefficients
if $D_B<D_A$ and larger if $D_B>D_A$.
The results displayed in Fig. \ref{figXXXX} prove that, for each description method, the width in a concentrated system is larger than the width in a dilute system
if $D_B<D_A$ and smaller if $D_B>D_A$. Hence, in the entire range of ratios of diffusion coefficients and for deterministic
as well as stochastic methods, the width in a concentrated system
is closer to the width obtained for identical diffusion coefficients. 
As for the front speed, the departure from the dilution limit reduces the effects induced by the difference between the diffusion coefficients.

\section{Conclusion}
We have performed kinetic Monte Carlo simulations of the master equation associated with a chemical system involving two species A and B.
The two species have two different diffusion coefficients, $D_A$ and $D_B$, and are engaged in the autocatalytic reaction $\rm{A}+\rm{B} \rightarrow 2\rm{A}$.
The effects of fluctuations on the FKPP wave front have been studied in the cases of a dilute solution
and a concentrated solution in which cross-diffusion cannot be neglected. 

In the case of a dilute system, the linearization of the deterministic equations with a cutoff in the leading edge of the front
leads to a speed shift independent of the diffusion coefficient $D_B$ of the consumed species. The speed shift obtained for two different diffusion coefficients
is the same as in the case $D_A=D_B$.
The main result deduced from the master equation is that
the front speed sensitively depends on the diffusion coefficient $D_B$. For $D_B$ larger than $D_A$,
the front speed decreases as $D_B$ increases and is significantly smaller than the prediction of the linear cutoff theory.
The speed decrease obtained for large values of $D_B/D_A$ is related to the number $N_{B_\varepsilon}$ of B particles
at the position of the most advanced A particle in the leading edge of the front.
When species B diffuses faster that species A, $N_{B_\varepsilon}$ is significantly smaller than the steady-state value $N_0$.\\

We carefully derived the nontrivial expression of the master equation in a concentrated system with cross-diffusion.
The transition rates are deduced from the diffusion fluxes in the linear domain of irreversible thermodynamics.
The transition rates associated with diffusion depend on the number of particles not only in the departure cell but also in the arrival cell.
Qualitatively, the conclusions drawn for a dilute solution and $D_A \neq D_B$ remain valid, but the front properties deduced from the master equation with cross-diffusion
depart less from those obtained for $D_A=D_B$. The dependence of the front properties on $D_B/D_A$ in a concentrated system are softened 
with respect to the dilute case. Cross-diffusion mitigates the impact of the difference between the diffusion coefficients.

\section{Acknowledgments}
This publication is part of a project that has received
funding from the European Union’s Horizon 2020 (H2020-EU.1.3.4.) research and innovation program under the Marie
Sklodowska-Curie Actions (MSCA-COFUND ID 711859) and from the Polish Ministry of Science and Higher Education
for the implementation of an international cofinanced project.

\end{document}